\documentclass[twocolumn,pra,superscriptaddress,showpacs,aps]{revtex4-1}
\usepackage{graphicx} 
\usepackage{dcolumn}
\usepackage{bm}
\usepackage{graphicx}
\usepackage{times}
\usepackage{soul}
\usepackage{color}
\usepackage{braket}
\usepackage{natbib}
\usepackage{floatrow}
\usepackage{physics}
\usepackage{comment}
\usepackage[normalem]{ulem}

\usepackage{color}
\usepackage{soul,xcolor}
\usepackage[normalem]{ulem}
\definecolor{black}{rgb}{0,0,0}
\definecolor{blue}{rgb}{0,0,1}
\definecolor{green}{rgb}{0,1,0}
\definecolor{red}{rgb}{1,0,0}
\definecolor{brown}{rgb}{0.4,0.2,0}
\definecolor{darkgreen}{rgb}{0,0.7,0}
\definecolor{darkblue}{rgb}{0.0,0.0,0.5}
\definecolor{red}{rgb}{1,0,0}
\definecolor{deepmagenta}{rgb}{0.8, 0.0, 0.8}

\usepackage[colorlinks,
citecolor = blue,
linkcolor = blue,
urlcolor  = blue,
anchorcolor = blue,
breaklinks = true
]{hyperref}

\def \MPQ{Max Planck Institute of Quantum Optics, 85748 Garching, Germany}

\def \IST{Institute of Science and Technology Austria, Am Campus 1, 3400 Klosterneuburg, Austria}
\def \ITP{CAS Key Laboratory of Theoretical Physics, Institute of Theoretical Physics,
	Chinese Academy of Sciences, Beijing 100190, China}
\def \IBK{Institute for Theoretical Physics, University of Innsbruck, Innsbruck A-6020, Austria}
\def \IQOQI{Institute for Quantum Optics and Quantum Information,
	Austrian Academy of Sciences, Innsbruck A-6020, Austria}
\def \Heidelberg{Institute for Theoretical Physics, Heidelberg University, Philosophenweg 16, 69120 Heidelberg, Germany}

\def \CAStopo{CAS Center for Excellence in Topological Quantum Computation \& School of Physical Sciences, \\ University of Chinese Academy of Sciences, Beijing 100049, China}

\begin{document}
	
	\title{Variational theory of angulons and their rotational spectroscopy}
	\author{Zhongda Zeng}\affiliation{\MPQ}\affiliation{\IBK}\affiliation{\IQOQI}
	\author{Enderalp Yakaboylu}\affiliation{\MPQ}
	\author{Mikhail Lemeshko}\affiliation{\IST}
	\author{Tao Shi}\affiliation{\ITP}\affiliation{\CAStopo}
	\author{Richard Schmidt}\affiliation{\MPQ}\affiliation{\Heidelberg}
	\preprint{}
	
	\begin{abstract}
		The angulon, a quasiparticle formed by a quantum rotor dressed by the excitations of a many-body bath, can be used to describe an impurity rotating in a fluid or solid environment. 	Here we propose a coherent state ansatz in the co-rotating frame which provides a comprehensive theoretical description of angulons. We reveal the quasiparticle properties, such as energies, quasiparticle weights and spectral functions, and show that our ansatz yields a persistent decrease in the impurity's rotational constant due to many-body dressing, consistent with experimental observations. From our study, a picture of the angulon emerges as an effective spin interacting with a magnetic field that is self-consistently generated by the molecule's rotation. Moreover, we discuss rotational spectroscopy, which focuses on the response of rotating molecules to a laser perturbation in the linear response regime.  Importantly, we take into account initial-state interactions that have been neglected in prior studies and reveal their impact on the excitation spectrum. To examine the angulon instability regime, we use a single-excitation ansatz and obtain results consistent with experiments, in which a broadening of spectral lines is observed while phonon wings remain highly suppressed due to initial-state interactions.
	\end{abstract}
	
	\date{\today}
	\maketitle

	\section{Introduction}	

	The angulon is a polaron-like quasiparticle that is formed by a rotating quantum impurity dressed by many-body excitations~\cite{lemeshko_molecular_2017}.	Generally, polaron models provide efficient descriptions of complex quantum many-body systems; historically starting with the description of how electrons move through a solid-state lattice and become dressed by lattice distortions thereby forming polaron quasiparticles~\cite{landau_electron_1933,Pekar,frohlich_electrons_1954,devreese_frohlich_2015}. Likewise, the angulon model considers a rotating impurity -- a quantum rotor -- dressed by phonon (or other kinds of) excitations  carrying angular momentum~\cite{schmidt_rotation_2015,schmidt_deformation_2016}. 	 
	
 	 One application of angulon theory are molecules embedded in superfluid helium nanodroplets.  This system has attracted great interest in molecular physics and chemistry in recent decades \cite{toennies_superfluid_2004,yang_helium_2012,PhysRevLett.110.093002}. Here the nanodroplets act as a stable and efficient refrigerator, cooling molecules to a  temperature of $\mathord{\sim}0.38$ Kelvin. Acting as an isolating matrix these droplets also provide a clean environment to  study molecules using spectroscopy or observe their chemical reactivity \cite{hartmann_direct_1996,grebenev_superfluidity_1998,stienkemeier_spectroscopy_2006,callegari_helium_2001, LugovojJCP00}. While helium's superfluidity prevents collisional and Doppler broadening of molecular spectral lines, the interaction between the molecule and helium causes a shift and an sometimes anomalous broadening of spectral lines in rotational spectroscopy \cite{morrison_rotational_2013}. This molecule-superfluid system can be theoretically understood as an impurity with rotational degrees of freedom embedded in a many-body bath, and be studied by first principle calculations such as quantum Monte Carlo \cite{zillich_roton-rotation_2004,zillich_quantum_2004,zillich_rotational_2010}. The angulon picture offers a relatively simple, approximate description of this complex molecular problem and is to a large extent consistent with experimental results \cite{lemeshko_molecular_2017,cherepanov_fingerprints_2017, lemeshko_quasiparticle_2017, PhysRevLett.118.085302, CherepanovPRA21, CherepanovNJP22}. 
	 
	Apart from molecules in superfluids, angulon theory has been applied to study rotation of organic cations in hybrid metal halide perovskites~\cite{AngulonPerovskite22}, molecules  immersed in a Bose-Einstein-Condensate~\cite{WillPRA19, PhysRevA.94.041601}, as well as transfer of angular momentum between electrons and crystal lattice in solids~\cite{MentinkPRB19}.
	
	Despite its relative simplicity, the angulon model is still challenging to solve due to the infinite dimension of the Hilbert space of phonons and the non-Abelian nature of the $SO(3)$ rotation group of the rotor \cite{varshalovich1988quantum,PhysRevLett.121.165301,PhysRevB.96.085410}.  This renders prior variational theories incomplete \cite{schmidt_rotation_2015, schmidt_deformation_2016}. On the one hand, they are unable to describe the renormalization of rotational constants in a fully general way. In particular, previous variational theories consistently predict a surprising result of an increase in rotational constants in the low-to-intermediate density regime. On the other hand, angulon theory predicts significant phonon wings, which dominate the spectrum in instability regimes. However, such pronounced phonon wings are hardly seen in experiments.
	
	The goal of the present paper is to resolve the two issues mentioned above. Concerning the first challenge, we propose a coherent state ansatz in the co-rotating frame and conduct a variational study of angulons. We find that the ground state can be described by a macroscopic wavefunction that is a product state of a bosonic and anomalous spin coherent state. From this ansatz, a simple picture of angulons arises, which can be used to access both static and dynamic properties. Here angulons present effective spin degrees of freedom interacting with a self-generated magnetic field. The resulting renormalized rotational constant is always decreased in agreement with physical intuition.
	
	Concerning the second challenge, the phonon wings,  we study the rotational spectroscopy of the $L=0\rightarrow1$ transition within the linear response theory using a single-excitation ansatz. Importantly, in contrast to previous works \cite{schmidt_rotation_2015}, we take fully into account the interactions in the initial state of the problem.	We show that while an instability regime in the spectrum persists, phonon wings become highly suppressed which is consistent with experiments. This emphasizes a significant role of the molecule-bath initial-state interaction.
	
	The article is structured as follows. In Sec.~\ref{subsec:Angulon}, the angulon model is revisited and the single-excitation ansatz is briefly introduced. In Sec.~\ref{subsec:Ansatz}, we present the coherent state ansatz in the rotor's co-rotating frame, and examine static properties such as the renormalization of the rotational constant (Sec.~\ref{subsec:Ground}). In addition, real-time evolution is employed to study the quasiparticle spectrum (Sec.~\ref{subsec:rte}).  Sec.~\ref{sec:spectroscopy} is concerned with a detailed study of the rotational spectra. We find that numerical predictions based on the single-excitation ansatz are compatible with experiments, exhibiting an instability regime but no phonon wings if one considers an equilibrium initial-state within the linear response theory. In Sec~\ref{sec:conclusion}, we summarize this work and discuss potential generalizations.

	\section{Coherent-state Angulon}\label{sec:CoherentAngulon}
	\subsection{Angulon model}\label{subsec:Angulon}
	We consider a linear molecule immersed in a weakly-interacting superfluid environment at zero temperature. The system can be described by a quantum rotor dressed by the Bogoliubov phonons excited from a weakly interacting bosonic bath \cite{landau2013quantum,bogolyubov_theory_1947,pitaevskii_bose-einstein_2016}. The Hamiltonian is given by \cite{schmidt_rotation_2015}: 
	\begin{equation}
	\label{eq:Ham}
		\begin{split}\hat{H}= & B\boldsymbol{\hat{J}}^{2}+\sum_{k\lambda\mu}\omega_{k}\hat{b}_{k\lambda\mu}^{\dagger}\hat{b}_{k\lambda\mu} 
			\\
			+&\sum_{k\lambda\mu}U_{\lambda}(k)[Y_{\lambda\mu}^{*}(\hat{\theta},\hat{\phi})\hat{b}_{k\lambda\mu}^{\dagger}+Y_{\lambda\mu}(\hat{\theta},\hat{\phi})\hat{b}_{k\lambda\mu}],\end{split}
	\end{equation}
	where $\hbar\equiv1$ and $\sum_{k}\equiv\int dk$. We note that the model is not expected to accurately describe molecules in strongly interacting superfluids, such as $^4$He, starting from first principles. However, much qualitative insights on molecular rotations in various environments can be gathered by treating the model Hamiltonian phenomenologically. The Hamiltonian consists of three terms. The first term represents the rotational kinetic energy of a rotor where $B$ is the rotational constant and $\boldsymbol{\hat{J}}$ is the angular momentum operator in the laboratory frame. The second term represents the kinetic energy of the phonons. The bosonic operators, $\hat{b}_{k\lambda\mu}^{(\dagger)}$, are given in the angular momentum representation, where $k=|\boldsymbol{k}|$ indicates the momentum magnitude; $\lambda$ and $\mu$ label the angular momentum quantum number and its projection onto the $z$ axis in the lab frame, respectively. Here we approximate the superfluid bath with phonons with dispersion relation $\omega_{k}=\sqrt{\epsilon_{k}(\epsilon_{k}+2g_{\text{bb}}n)}$, where $g_{bb}=4\pi a_{\text{bb}}/m$. We set the boson-boson scattering length to $a_{\text{bb}}=3.3(mB)^{-1/2}$ and the rotational constant to $B= 2\pi \times 1$~GHz.
	
	The last term of Eq.~\eqref{eq:Ham} represents the interaction between the rotor and the phonon bath, which couples the angles of the rotor, $\hat{\theta}$ and $\hat{\phi}$, with the phonon fluctuations; $j_{\lambda}(kr)$ are the spherical Bessel functions of the first kind and $Y_{\lambda\mu}(\hat{\theta},\hat{\phi})$ are the spherical harmonics. The interaction strength is given by $U_{\lambda}(k)=u_{\lambda}[\frac{8nk^{2}\epsilon_{k}}{\omega_{k}(2\lambda+1)}]^{1/2}\int drr^{2}f_{\lambda}(r)j_{\lambda}(kr)$, where we consider Gaussian form factors, $f_{\lambda}(r)=(2\pi)^{-3/2}e^{-r^{2}/(2r_{\lambda}^{2})}$. The interaction amplitudes and ranges are chosen as in previous works \cite{schmidt_rotation_2015} to allow for direct comparison: $u_{0}=1.75u_{1}=218B$ and $r_{0}=r_{1}=1.5(mB)^{-1/2}$, respectively. Apart from the rotor, we  introduce the angular momentum operators of the phonon bath $\hat{\Lambda}^{\alpha}=\sum_{k\lambda\mu\nu}\hat{b}_{k\lambda\mu}^{\dagger}\sigma_{\mu\nu}^{\lambda,\alpha}\hat{b}_{k\lambda\nu}$, and further introduce the total angular momentum $\boldsymbol{\hat{L}}=\boldsymbol{\hat{J}}+\hat{\boldsymbol{\Lambda}}$. It is easy to check that the angulon Hamiltonian commutes with $\mathbf{\hat{L}}^{2}$ and $\hat{L}_{z}$, which shows that the eigenstates can be labeled by the two quantum numbers $L$ and $M$.
	
	Several studies of the angulon problems have been based on a single-excitation ansatz \cite{schmidt_rotation_2015,PhysRevA.94.041601,PhysRevLett.118.085302,cherepanov_fingerprints_2017,lemeshko_quasiparticle_2017}, which is given by 
	\begin{equation}\label{eq:single_excitation_ansatz}
		\begin{split}|\psi_{LM}\rangle= & Z^{1/2}|0\rangle|LM\rangle \\
			+&\sum_{k\lambda\mu}\sum_{jn}\beta_{k\lambda j}C_{jn,\lambda\mu}^{LM}\hat{b}_{k\lambda\mu}^{\dagger}|0\rangle|jn\rangle.
		\end{split}
	\end{equation}
	The first term indicates the non-interacting vacuum state. Here $|0\rangle$ labels the bosonic vacuum, $|LM\rangle$ labels the rotor state, and $Z$ is the quasiparticle renormalization factor satisfying the normalization condition, $|Z|+\sum_{k\lambda j}|\beta_{k\lambda j}|^{2}=1$. The second term in Eq.~\eqref{eq:single_excitation_ansatz} indicates the single-excitation state in which the Clebsch-Gordan coefficients $C_{jn,\lambda\mu}^{LM}$ incorporate the total angular momentum conservation of the rotor and excited phonons. The wavefunction is labeled by two quantum numbers, the total angular momentum quantum number $L$ and its projection onto the $z$ axis $M$. 
	
	This ansatz, coinciding with a generalized second-order perturbation theory \cite{lemeshko_molecular_2017}, successfully explains the anomalous broadening of spectral lines in spectroscopy experiments \cite{cherepanov_fingerprints_2017}, and the renormalization of rotational constants at weak coupling, e.g., small $u_0/B$  \cite{lemeshko_quasiparticle_2017}. However, it fails to describe the renormalization of the rotational constants in the intermediate-density regime, which exhibits quasiparticle instability. In the following, we will propose a coherent state ansatz in the co-rotating frame that resolves this issue and compare it with the single-excitation ansatz.

	\subsection{Variational ansatz}\label{subsec:Ansatz}
	
	In the Fr\"ohlich polaron \cite{frohlich_electrons_1954}, the Lee-Low-Pines (LLP) transformation can be used to decouple the impurity's degree of freedom from the many-body bath \cite{PhysRev.90.297}. Then the remaining bosonic model can be approximately solved with a coherent state \cite{PhysRevLett.117.113002} or Gaussian state ansatz \cite{PhysRevA.93.043606}. The overall variational wavefunction can be represented as a product state ansatz between the impurity and the bath coupled by the canonical transformation. In other words, the transformation entangles   the two parts. This method thereby goes beyond the mean-field framework, representing a generalized mean-field theory that is a first step towards recently-developed non-Gaussian state methods \cite{shi_variational_2018}.
	
	The LLP transformation is a \textit{translational} transformation to the co-moving frame of the impurity that results from the total momentum conservation of the whole system. In the angulon system, similarly, the total angular momentum square $\mathbf{\hat{L}}^2$ and projection $\hat{L}_z$ are conserved. Analogous to the LLP transformation, the problem can be simplified by a \textit{rotational} transformation to the co-rotating frame, as shown in Ref.~\cite{schmidt_deformation_2016}: 
	\begin{equation}
		\hat{S}=e^{-i\hat{\phi}\otimes\hat{\Lambda}^{z}}e^{-i\hat{\theta}\otimes\hat{\Lambda}^{y}}e^{-i\hat{\gamma}\otimes\hat{\Lambda}^{z}}.
	\end{equation}
	The Hamiltonian in the rotating frame reads
	\begin{equation}\label{eq:Hamiltonian_transfer}
		\begin{split}\hat{\mathcal{H}}= & \hat{S}^{-1}\hat{H}\hat{S}=B(\hat{\boldsymbol{J^{\prime}}}-\hat{\boldsymbol{\Lambda}})^{2}\\
			& +\sum_{k\lambda\mu}\omega_{k}\hat{b}_{k\lambda\mu}^{\dagger}\hat{b}_{k\lambda\mu}+\sum_{k\lambda}V_{\lambda}(k)[\hat{b}_{k\lambda0}^{\dagger}+\hat{b}_{k\lambda0}],
		\end{split}
	\end{equation}
	where $V_{\lambda}(k)=\sqrt{(2\lambda+1)/4\pi}U_{\lambda}(k)$.	Here $\hat{\boldsymbol{J^{\prime}}}$ denotes the anomalous angular momentum operator, which represents the total angular momentum of the system and satisfies the anomalous commutation relations:
	\begin{equation}\label{eq:commutation_relation}
		[\hat{J}^{\prime\alpha},\hat{J}^{\prime\beta}]=-i\epsilon_{\alpha\beta\gamma}\hat{J}^{\prime\gamma}.
	\end{equation}
	Here the indices refer to $x,y,z$ (coordinates in the rotating frame of the molecule) and $\epsilon_{\alpha\beta\gamma}$ is the Levi-Civita symbol.
	
	The angular state is characterized by three quantum numbers, $L$, $M$ and $n$, corresponding to the eigenvalues of the operators, $\mathbf{\hat{J}}^2$, $\hat{J}^z$, and $\hat{J^{\prime}}^z$, where  $\hat{\mathbf{J^\prime}}^{2}= \mathbf{\hat{J}}^2$. Note that the system conserves the square of angular momentum and the angular momentum projection onto the lab frame, but not onto the rotating frame.
	
	In the slowly-rotating limit, $B\rightarrow0$, the~Hamiltonian reduces to a purely bosonic one and can be diagonalized exactly by a displacement operator, $\hat{U}= \exp[-\sum_{k\lambda\mu} \frac{V_{\lambda}(k)}{\omega_{k}}(\hat{b}_{k\lambda0}^{\dagger}-\hat{b}_{k\lambda0})]$. The ground state is thus a coherent state which contains an infinite number of phonon excitations, and the ground-state energy simply reads $E_{0}=-\sum_{k\lambda}V_{\lambda}^{2}(k)/\omega_{k}$.
	
	Based on the above discussion, we propose the following variational ansatz in the co-rotating frame: 
	\begin{equation}\label{eq:coherent}
		|\psi\rangle=\sum_{n}g_{n}|LMn\rangle\otimes|C\rangle,
	\end{equation}
	which is a product state between the angular state describing the total angular momentum, and the bosonic coherent state describing the superfluid bath: 
	\begin{equation}
		|C\rangle=\exp(\sum_{k\lambda\mu}\hat{b}_{k\lambda\mu}^{\dagger}\beta_{k\lambda\mu}-\hat{b}_{k\lambda\mu}\beta_{k\lambda\mu}^{*})|0\rangle.
	\end{equation}
	Within a given $L$ sector, one can substitute the operator $\hat{\mathbf{J^{\prime}}}^{2}$ with its eigenvalue $L(L+1)$. Due to the non-commutative structure of $\hat{J^{\prime}}^{\alpha}$, we consider a superposition in the $n$ channel represented by the variational parameters $g_n$. For the coherent bath, the $\beta_{k\lambda\mu}$ are variational parameters which are optimized by minimizing the variational energy. In dynamical problems, they are promoted to time-dependent variables. We would like to emphasize that although the ansatz appears like a mean-field theory in the transformed frame, the overall ansatz in the lab frame, $\hat{S}|\psi\rangle$, includes entanglement between the rotor and the bath through the canonical transformation $\hat{S}$, and thus goes beyond the mean-field framework.
	
	Although a coherent state ansatz has been discussed in Ref.~\cite{lemeshko_quasiparticle_2017,schmidt_deformation_2016}, it has not been yet considered as a variational state.	In Ref.~\cite{lemeshko_quasiparticle_2017}, it was shown to yield a phenomenological prediction of the renormalization of the rotational constants in the strong-coupling regime. In contrast, Ref.~\cite{schmidt_deformation_2016} considered a single-excitation ansatz on top of a coherent bath, which reveals a critical density beyond which the impurity acquires one quantum of angular momentum from the many-particle bath. In these two studies, the displacement vector $\beta$ is set as fixed by the solution for the $L=0$ sector, where the rotor and the phonon cloud do not rotate. In the present work, the displacement operator is generalized to include variational parameters. As a result, we can fully take into  account the phonon cloud deformation caused by the rotor's rotation. Moreover, it is straightforward to extend the study to real-time evolution and dynamical problems.

	\begin{figure*}
		\centering    
		\includegraphics[width=1.0\columnwidth]{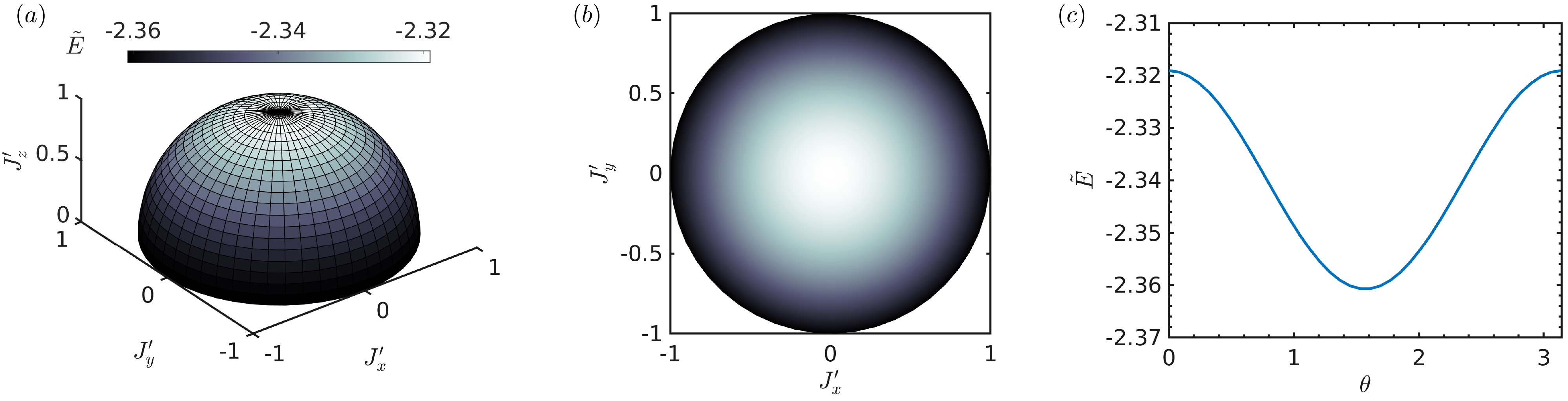}
		\caption{(a) Variational energy $\tilde{E}=E/B$ on the surface of the effective spin Bloch sphere for $L=1$. The Bloch sphere 	is constructed by the vector $\boldsymbol{J^{\prime}}$ parametrized by the polar angles $(\theta,\phi)$. (b) Projection of the variational energy into the $J_{x}^{\prime}-J_{y}^{\prime}$ plane. The energy is independent of $\phi$ and rotationally invariant about the $J_{z}^{\prime}$ axis.  (c) Variational energy as a function of $\theta$, which is minimized at the $\theta=\pi/2$. }
		\label{fig:BlochSphere}
	\end{figure*}
	
	Using Eq.~\eqref{eq:coherent}, the variational energy reads: 
	\begin{equation}
		\begin{split}E= & \langle\psi|\hat{\mathcal{H}}|\psi\rangle=BL(L+1)-2B\boldsymbol{J^{\prime}}\cdot\boldsymbol{\Lambda}+B\boldsymbol{\Lambda}\cdot\boldsymbol{\Lambda}\\
			& +\sum_{k\lambda\mu}W_{k\lambda}\beta_{k\lambda\mu}^{*}\beta_{k\lambda\mu}+\sum_{k\lambda}V_{\lambda}(k)(\beta_{k\lambda0}^{*}+\beta_{k\lambda0})
		\end{split}
	\end{equation}
	where 
	\begin{equation}
		W_{k\lambda}\equiv\omega_{k}+B\lambda(\lambda+1),
	\end{equation}
	and the \textit{average} quantities of angular momentum, $\boldsymbol{\Lambda}\equiv\langle\psi|\hat{\boldsymbol{\Lambda}}|\psi\rangle$ and  $\boldsymbol{J^{\prime}}\equiv\langle\psi|\hat{\boldsymbol{J^{\prime}}}|\psi\rangle$, are given by 
	\begin{equation}
		\begin{split}\Lambda^{\alpha}= & \sum_{k\lambda\mu\nu}\beta_{k\lambda\mu}^{*}\sigma_{\mu\nu}^{\lambda,\alpha}\beta_{k\lambda\nu},\\
			J^{\prime\alpha}= & \sum_{nn^{\prime}}g_{n}^{*}g_{n^{\prime}}\langle LMn|\hat{J}^{\prime\alpha}|LMn^{\prime}\rangle.
		\end{split}
	\end{equation}

	\subsection{Ground-state properties}\label{subsec:Ground}
	
	As a result of the product-state structure in Eq.~\eqref{eq:coherent}, the angular and coherent states variational parameters can be optimized separately in an iterative way where the respective other part is kept fixed \cite{wang_zero-temperature_2019}.  In what follows, we  illustrate the optimization scheme in detail.
	
	\textbf{\textit{Angular state.}} For the angular state, the effective Hamiltonian is obtained by tracing out the bosonic bath: 
	\begin{equation}\label{eq:H_r}
		\begin{split}\hat{\mathcal{H}}_{\text{rot}}\equiv & \langle C|\hat{\mathcal{H}}|C\rangle=-2B\boldsymbol{\Lambda}\cdot\hat{\boldsymbol{J^{\prime}}}+f(\beta,\beta^{*}),\end{split}
	\end{equation}
	with 
	\begin{equation}
		\begin{split} & f(\beta,\beta^{*})=BL(L+1)+B\boldsymbol{\Lambda}\cdot\boldsymbol{\Lambda}\\
			& +\sum_{k\lambda\mu}W_{k\lambda}\beta_{k\lambda\mu}^{*}\beta_{k\lambda\mu}+\sum_{k\lambda}V_{\lambda}(k)(\beta_{k\lambda0}^{*}+\beta_{k\lambda0}).
		\end{split}
	\end{equation}
	Remarkably, Eq.~\eqref{eq:H_r} implies that the effective model reduces to a single ``anomalous'' (in the sense of commutation relations) high-dimensional spin in an effective external field, $\boldsymbol{B}_{\text{eff}}=2B\boldsymbol{\Lambda}$, that emerges from the collective rotation of the phonon bath. Here the anomalous spin satisfies the anomalous commutation relations. The effective field, which is real, can be parameterized by its amplitude and two polar angles, $\boldsymbol{B}_{\text{eff}}=2B(\Lambda^{x},\Lambda^{y},\Lambda^{z})=|\boldsymbol{B}_{\text{eff}}|(\sin\theta\cos\phi,\sin\theta\sin\phi,\cos\theta)$. We note that the form of the effective magnetic field is a reminiscent of a magnetic monopole, as it is solely along the radial direction. As a matter of fact, it has been previously demonstrated that the angulon can be seen as a point charge on a 2-sphere interacting with a magnetic monopole \cite{PhysRevLett.119.235301}.
	
	The Hamiltonian Eq.~\eqref{eq:H_r} can be represented in a matrix form and solved numerically. One finds that the energy is minimized when the anomalous spin aligns with the effective field, similarly to the conventional single spin in a real magnetic field. One can thus employ a rotational transformation generated by the anomalous angular momentum $\hat{\boldsymbol{J^{\prime}}}$ to diagonalize the Hamiltonian, where the transformation is given by 
	\begin{equation}
		\hat{D}^{\prime}(\alpha,\beta,\gamma)=e^{-i\alpha\hat{J}^{\prime z}}e^{-i\beta\hat{J}^{\prime y}}e^{-i\gamma\hat{J}^{\prime z}},
	\end{equation}
	which corresponds to a left-handed rotational transformation. The Hamiltonian after transformation is given by $\hat{D}^{\prime\dagger}(-\phi,-\theta,0)\hat{\mathcal{H}}_{\text{rot}}\hat{D}^{\prime\dagger}(-\phi,-\theta,0)=-\hat{J}^{\prime z}$ and the corresponding ground state  is given by
	\begin{equation}
		\hat{D}^{\prime}(-\phi,-\theta,0)|LML\rangle = \sum_{n}g_{n}|LMn\rangle,
	\end{equation} 
	where $D_{nm}^{\prime}$ is the Wigner D-matrix. To avoid confusion with the normal spin coherent state, we call this state the \textit{anomalous spin coherent state}.
	
	Following Schwinger's oscillator method \cite{sakurai2011modern,schwinger1965quantum}, we derive an analytical expression for the anomalous spin coherent state, whose superposition coefficients are given by 
	\begin{equation}\label{eq:spinCoherentState}
		g_{n}=\begin{pmatrix}2L\\
			L+n
		\end{pmatrix}^{1/2}(\cos\frac{\theta}{2})^{L+n}(\sin\frac{\theta}{2})^{L-n}e^{-i\phi(L-n)},
	\end{equation}
	which is similar to the normal spin coherent state up to a phase.
	
	The anomalous spin coherent states can be characterized by the expectation values of the angular momentum, $\boldsymbol{J^{\prime}}(\theta,\phi)=L(\sin\theta\cos\phi,\sin\theta\sin\phi,\cos\theta)$, pointing at the surface of the Bloch sphere as shown in Fig.~\ref{fig:BlochSphere}(a). These states are macroscopic quantum states which minimize the variance of the angular momentum. Remarkably, it thus turns out that the angulon ground state is described by a product state of the bosonic coherent states and the anomalous spin coherent states.

	\textbf{\textit{Bosonic state.}} For the coherent bath one can optimize the variational parameters by imaginary-time evolution,
	\begin{equation}
		\partial_{\tau}|C\rangle=-(\hat{\mathcal{H}}_{b}-E)|C\rangle,
	\end{equation}
	where $\hat{\mathcal{H}}_{b}=\sum_{nn^{\prime}}g_{n}^{*}g_{n^{\prime}}\langle LMn|\hat{\mathcal{H}}|LMn^{\prime}\rangle$ is the effective Hamiltonian for bosons derived by tracing out the angular states. Correspondingly, one can derive the equation of motion for the variational parameters, 
	\begin{equation}\label{eq:coherent_ite}
		\partial_{\tau}\beta_{k\lambda\mu}=-\eta_{k\lambda\mu},
	\end{equation}
	with an effective mean-field Hamiltonian 
	\begin{equation}
		\begin{split}\eta_{k\lambda\mu}= & W_{k\lambda}\beta_{k\lambda\mu}+\delta_{\mu0}V_{\lambda}(k)\\
			& +2B(\boldsymbol{\Lambda}-\boldsymbol{J^{\prime}})\cdot\sum_{\nu}\boldsymbol{\sigma}_{\mu\nu}^{\lambda}\beta_{k\lambda\nu}.
		\end{split}
		\label{eq:mean_field_eta}
	\end{equation}
	The variational energy converges to a local minimum for sufficiently long evolution time, which is a saddle point of the mean-field Hamiltonian, i.e. $\eta_{k\lambda\mu}\overset{!}{=}0$.
	
	Above we described a method to approach the local energy minimum for both angular and bosonic states by iteratively fixing the other. In this scheme, the angular momentum of bosons, $\boldsymbol{\Lambda}$, will evolve continuously in the imaginary-time evolution, while the exact diagonalization of angular states can induce $\boldsymbol{J^{\prime}}$ to have a sudden jump on the Bloch sphere. Finally, a global minimum will be reached when both processes are performed iteratively.
	
	An alternative approach, leading eventually to the same results, is to find the saddle point in a self-consistent way. In Fig.~\ref{fig:BlochSphere}, we scan the energies obtained by fixing the angular momentum $\boldsymbol{J^{\prime}}$ as given by the polar angles $(\theta,\phi)$, and performing imaginary-time evolution for the bosonic states. 
	The energies are always minimized when $J_{z}^{\prime}=0$, correspondingly $\theta=\pi/2$, due to the rotational symmetry along the $z^\prime$ axis. Moreover, by symmetry, $\phi$ is an irrelevant parameter for the static problem. Hence without loss of generality, we may set $\theta=\pi/2$ and $\phi=0$ in the following ground-state calculation.

	\begin{figure}[t]
		\centering  
		\includegraphics[width=1.0\columnwidth]{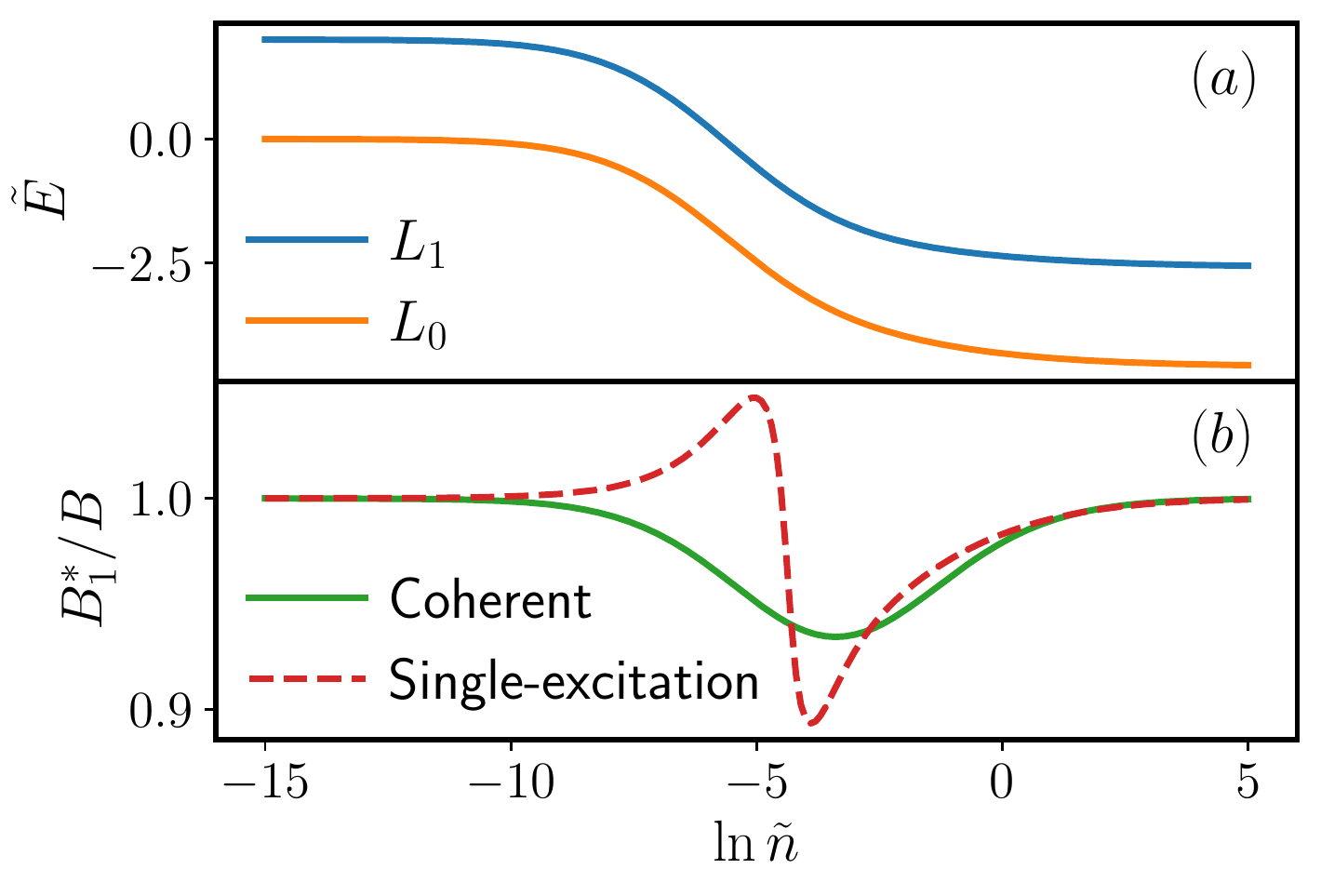}
		\caption{(a) Angulon energies $\tilde{E}$ for $L=0,1$ obtained by the coherent state ansatz. (b) Effective rotational constants $B^{*}$ obtained by the coherent state ansatz and single-excitation ansatz as a function of the dimensionless superfluid density $\tilde{n}=n(mB)^{-3/2}$. The results from the two Ansätze are consistent in the large-density regime, but the single-excitation ansatz predicts an increasing effective rotational constant in the low-to-intermediate density regime. In contrast the coherent state ansatz consistently predicts a decreasing rotational constant.}
		\label{fig:Energy}
	\end{figure}

	We next consider the bosonic states in this alternative approach. In addition to solving the ordinary differential equations for imaginary-time evolution, the saddle point can be found in a self-consistent manner. As discussed above, for the ground state of the $L=0$ sector the total angular momentum vanishes, $\boldsymbol{J}^{\prime}=0$. Then a saddle point solution of the bosonic states follows: 
	\begin{equation}
		\beta_{k\lambda\mu}^{(0)}=-\delta_{\mu0}\frac{V_{\lambda}(k)}{W_{k\lambda}},\label{eq:coherent_ground_wavefunction}
	\end{equation}
	with the corresponding ground-state energy: 
	\begin{equation}
		E_{0}=-\sum_{k\lambda}\frac{V_{\lambda}^{2}(k)}{W_{k\lambda}}.\label{eq:coherent_ground_energy}
	\end{equation}
	which has been referred to as deformation energy \cite{schmidt_deformation_2016}.

	\begin{figure}[t]
		\centering 
		\includegraphics[width=1.0\columnwidth]{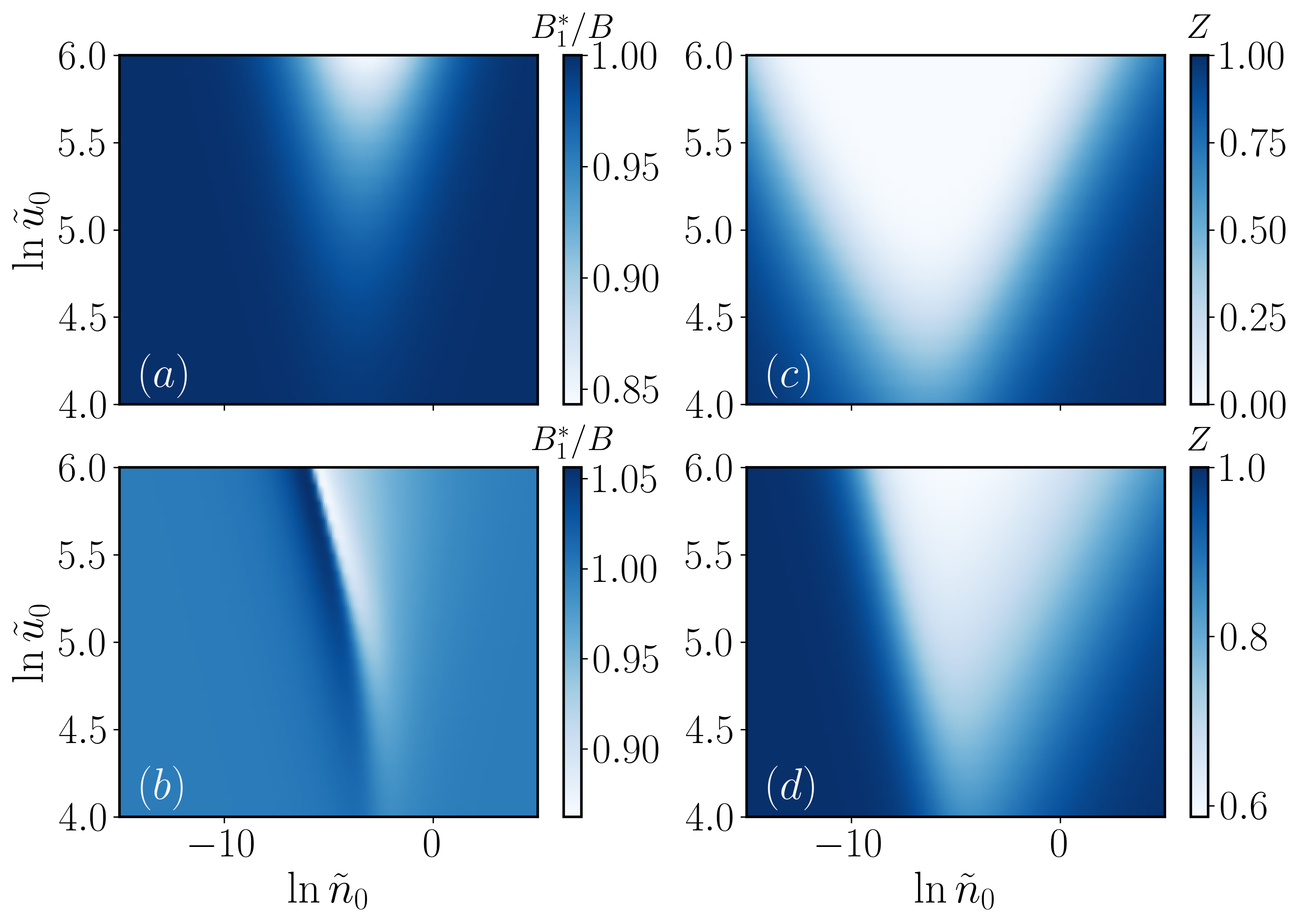}
		\caption{(a), (b) Effective rotational constants, and (c), (d) quasiparticle weights $Z$, obtained by the coherent state ansatz (top) and single-excitation ansatz (bottom) as a function of the superfluid density and the dimensionless rotor-boson interaction amplitude $\tilde{u}_{0}=u_{0}/B$.}
		\label{fig:BZ}
	\end{figure}

	\begin{figure*}
		\centering
		\includegraphics[width=1.0\columnwidth]{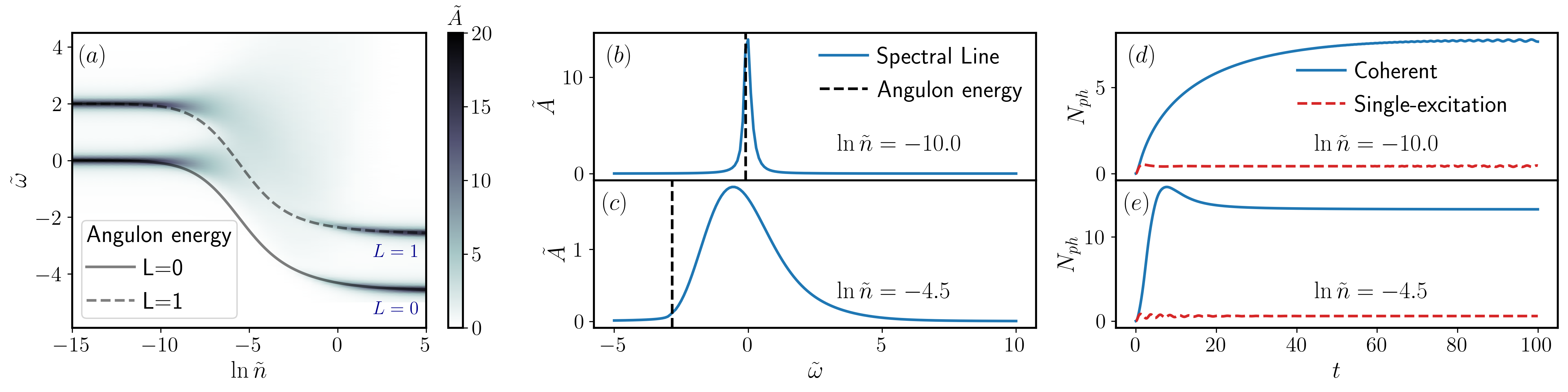}
		\caption{(a) Angulon spectral function for $L=0, 1$ as a function of the dimensionless superfluid density within the coherent state ansatz. The lines represent the angulon energies $\tilde{E}$ obtained by a saddle point analysis of the coherent state ansatz. Here we introduce the dimensionless frequency $\tilde{\omega}=\omega/B$. (b), (c) Cuts of the spectral function at a fixed density for $L=0$ in the low- and intermediate-density regime. (d), (e) Average number of phonon excitations in real-time evolution. The coherent state result (solid) is compared to a single-excitation ansatz (dashed). The long-time oscillation originates from the discretization of radial momentum $k$. }
		\label{fig:QuasiparticleSpectrum}
	\end{figure*}

	Also for general $L$ sectors, the $\lambda=0$ channel has a simple solution: 
	\begin{equation}
		\beta_{k00}=-\frac{V_{0}(k)}{\omega_{k}}.
	\end{equation}
	For the $\lambda=1$ channel, the variational parameters $\beta_{k\lambda\mu}$ are subject to an effective Hamiltonian, given by 
	\begin{equation}
		\begin{split} & \begin{pmatrix}W_{k1} & \sqrt{2}B(\Lambda^{x}-J^{\prime x}) & 0\\
				\sqrt{2}B(\Lambda^{x}-J^{\prime x}) & W_{k1} & \sqrt{2}B(\Lambda^{x}-J^{\prime x})\\
				0 & \sqrt{2}B(\Lambda^{x}-J^{\prime x}) & W_{k1}
			\end{pmatrix}\\
			& \times\begin{pmatrix}\beta_{k11}\\
				\beta_{k10}\\
				\beta_{k1-1}
			\end{pmatrix}=-\begin{pmatrix}0\\
				V_{1}(k)\\
				0
			\end{pmatrix}.
		\end{split}
		\label{eq:lambda1}
	\end{equation}
	Here we have taken $\Lambda^{y}=\Lambda^{z}=0$ since $\theta=\pi/2$ and $\phi=0$. From this one can derive the self-consistent equation: 
	\begin{equation}
		\Lambda^{x}=\sum_{k}\frac{4BV_{1}^{2}(k)W_{k1}(J^{\prime x}-\Lambda^{x})}{(W_{k1}^{2}-4B^{2}(J^{\prime x}-\Lambda^{x})^{2})^{2}}.\label{eq:selfconsistent}
	\end{equation}
	This equation can be solved numerically and one can then obtain the variational parameters $\beta_{k1\mu}$ by inserting $\Lambda^{x}$ back into Eq.~\eqref{eq:lambda1}.  The resulting energies are identical to the iterative procedure in imaginary-time evolution.

	\textbf{\textit{Renormalized rotational constant.}} The renormalization of the rotational constant is one of the key phenomena described by angulon theory. In analogy it is similar to how a phonon cloud leads to the renormalization of the electron's mass in the Fr\"ohlich model that describes the modified translational motion of a particle in a bosonic medium. In the angulon problem, the rotor excites a rotating phonon cloud and forms a quasiparticle. This deformation leads to a decreased effective rotational constant defined by 
	\begin{equation}
		B^{*}_{L}=\frac{E_{L}-E_{0}}{L(L+1)}.\label{eq:RotationalConstant}
	\end{equation}	
	
	As a benchmark for our variational solution, we show the energy of the angulon for the $L=0,1$ sectors in Fig.~\ref{fig:Energy}(a), and compare the effective rotational constants obtained by the coherent state ansatz and the single-excitation ansatz in Fig.~\ref{fig:Energy}(b). At large densities, the results from the two approaches are consistent.  However, in the low-density regime, the single-excitation ansatz predicts an increasing effective rotational constant. This would indicate the surprising result of a ``speeding up'' of the rotor which is both inconsistent with the physics of translational impurities as well as experimental observations of molecules in superfluid Helium nanodroplets. In contrast, the coherent state ansatz always predicts a decreasing rotational constant agreeing with the intuition that consistent dressing of a polaron cloud should hinder the rotation of the composite state of the rotor and its local environment.

	The quasiparticle weight $Z$ is an important quasiparticle property, characterizing how well-defined the quasiparticle is. It is defined as the absolute square of the overlap between the vacuum state and the ground state, 
	\begin{equation}
		Z=|\langle0|C\rangle|^{2}=e^{-N_{\text{ph}}},
	\end{equation}
	where $N_{\text{ph}}=\sum_{k\lambda\mu}|\beta_{k\lambda\mu}|^{2}$ is the particle number of phonons. When $Z$ is finite, the quasiparticle is well-defined, whereas $Z=0$ indicates the breakdown of the quasiparticle pictures akin to the ``orthogonality catastrophe'' described in fermionic systems \cite{anderson_infrared_1967}.
	In Fig.~\ref{fig:BZ}, we show the quasiparticle weight and effective rotational constants as function of the rotor-boson interaction amplitude and superfluid density. We find that compared to the single-excitation ansatz the quasiparticle weight is more suppressed by interactions in the coherent state ansatz due to its ability to describe the dressing of the rotor by a macroscopic number of phonons.

	\subsection{Spectral Function}\label{subsec:rte}
	We next consider the angulon spectral function obtained within the coherent state framework. The angulon Green's function is defined as $G(t)=\langle\psi(0)|\psi(t)\rangle$, where $|\psi(0)\rangle\equiv|LM0\rangle\otimes|0\rangle$ represents the unperturbed vacuum state, and $|\psi(t)\rangle=e^{-i\hat{\mathcal{H}}t}|\psi(0)\rangle$ indicates its time-evolution. The analytical structure of its Fourier transformation $G(\omega)$ in the complex frequency plane gives direct access to the angulon energy, lifetime, and quasi-particle weight. The quasiparticle spectral function is given by \cite{PhysRevX.2.041020, Schmidt_2018},
	\begin{equation}\label{eq.spectrum}
		A(\omega)=  2\text{Re}\int_{0}^{\infty}dte^{i\omega t}\langle\psi(0)|\psi(t)\rangle.
	\end{equation}
	Using the coherent state ansatz Eq.~\eqref{eq:coherent}, the spectral function is then given by
	\begin{equation}
		A(\omega)= 2\text{Re}\int_{0}^{\infty}dte^{i\omega t}g_{0}(t)e^{-\frac{1}{2}\sum_{k\lambda\mu}|\beta_{k\lambda\mu}(t)|^{2}},
	\end{equation}
	where the variational parameters are treated as time-dependent. The real-time evolution of $|\psi(t)\rangle$ is governed by the Schr\"odinger equation, $i\partial_{t}|\psi\rangle=\hat{H}|\psi\rangle$, from which one can derive the equations of motion for the variational parameters: 
	\begin{equation}\label{eq:rte_beta}
		\begin{split}\begin{split}i\partial_{t}\beta_{k\lambda\mu}=\end{split}
			& \delta_{\mu0}V_{\lambda}(k)+W_{k\lambda}\beta_{k\lambda\mu}\\
			& +2B(\boldsymbol{\Lambda}-\boldsymbol{J^{\prime}})\cdot\sum_{\nu}\boldsymbol{\sigma}_{\mu\nu}^{\lambda}\beta_{k\lambda\nu},
		\end{split}
	\end{equation}
	and 
	\begin{equation}\label{eq:rte_g}
		\begin{split}i\partial_{t}g_{n}= & g_{n}[BL(L+1)+2B\boldsymbol{J^{\prime}}\cdot\boldsymbol{\Lambda}-B\boldsymbol{\Lambda}\cdot\boldsymbol{\Lambda}\\
			& +\sum_{k\lambda}V_{\lambda}(k)\text{Re}\beta_{k\lambda0}]-2B\sum_{n^{\prime}}g_{n^{\prime}}\boldsymbol{J^{\prime}}_{nn^{\prime}}\cdot\boldsymbol{\Lambda},
		\end{split}
	\end{equation}
	where $\boldsymbol{J^{\prime}}_{nn^{\prime}}\equiv\langle LMn|\hat{\boldsymbol{J^{\prime}}}|LMn^{\prime}\rangle$.
	
	Importantly, as the initial state $|\psi(0)\rangle$ is a zero-angular momentum state of the bosons, in the coherent state evolution the boson angular momentum $\boldsymbol{\Lambda}$ as well as $\boldsymbol{J^{\prime}}$ remain zero (unlike for the ground state which acquires finite expectation values of these quantities). As a result, the equation of motion reduce to 
	\begin{equation}
		\begin{split}\begin{split}\end{split}
			& i\partial_{t}\beta_{k\lambda\mu}=\delta_{\mu0}V_{\lambda}(k)+W_{k\lambda}\beta_{k\lambda\mu},\\
			& i\partial_{t}g_{n}=g_{n}[BL(L+1)+\sum_{k\lambda}V_{\lambda}(k)\text{Re}\beta_{k\lambda0}].
		\end{split}
	\end{equation}
	Due to the simplicity of these equations, the time evolution can be solved analytically and one obtains the analytical expressions for the variational parameters: 
	\begin{equation}
		\beta_{k\lambda\mu}(t)=-\frac{\delta_{0\mu}V_{\lambda}(k)}{W_{k\lambda}}(1-e^{-iW_{k\lambda}t}),
	\end{equation}
	and 
	\begin{equation}
	    \begin{split}
		g_{n}(t)= & \delta_{0,n}\text{exp}\biggl(-iBL(L+1)t \\
		 & + i\sum_{k\lambda}\frac{V_{\lambda}^{2}(k)}{W_{k\lambda}}[1-\text{sinc}(W_{k\lambda}t)]t\biggl).
	    \end{split}
	\end{equation}

	Fig.~\ref{fig:QuasiparticleSpectrum} (a) shows the spectral function in dependence on the superfluid density. The spectral function is drastically broadened in the intermediate density regime even for the $L=0$ sector, which is consistent with the calculation of quasiparticle weights in Fig.~\ref{fig:BZ} (b). In Fig.~\ref{fig:QuasiparticleSpectrum} (b) and (c), we show cuts of the spectral function in the low- and intermediate-density regimes, respectively. The angulon spectral line is sharp in the former case and is red-shifted and broadened for intermediate densities. Intriguingly, such a drastic broadening has also been found for translational Bose polarons in the intermediate interaction regime \cite{PhysRevLett.117.113002} consistent with experimental observations \cite{PhysRevLett.117.055301,PhysRevLett.117.055302}.

	In Fig.~\ref{fig:QuasiparticleSpectrum} (d) and (e), we compare the time-dependent average number of phonon excitations obtained by the single-excitation and coherent state ansatz for two density regimes. In the single-excitation ansatz the excitation number is limited to one by construction, whereas the coherent state does not impose such a limit and it indeed shows a growth of the phonon number to values significantly above unity.
	
\begin{figure}
		\centering 
		\includegraphics[width=1.0\columnwidth]{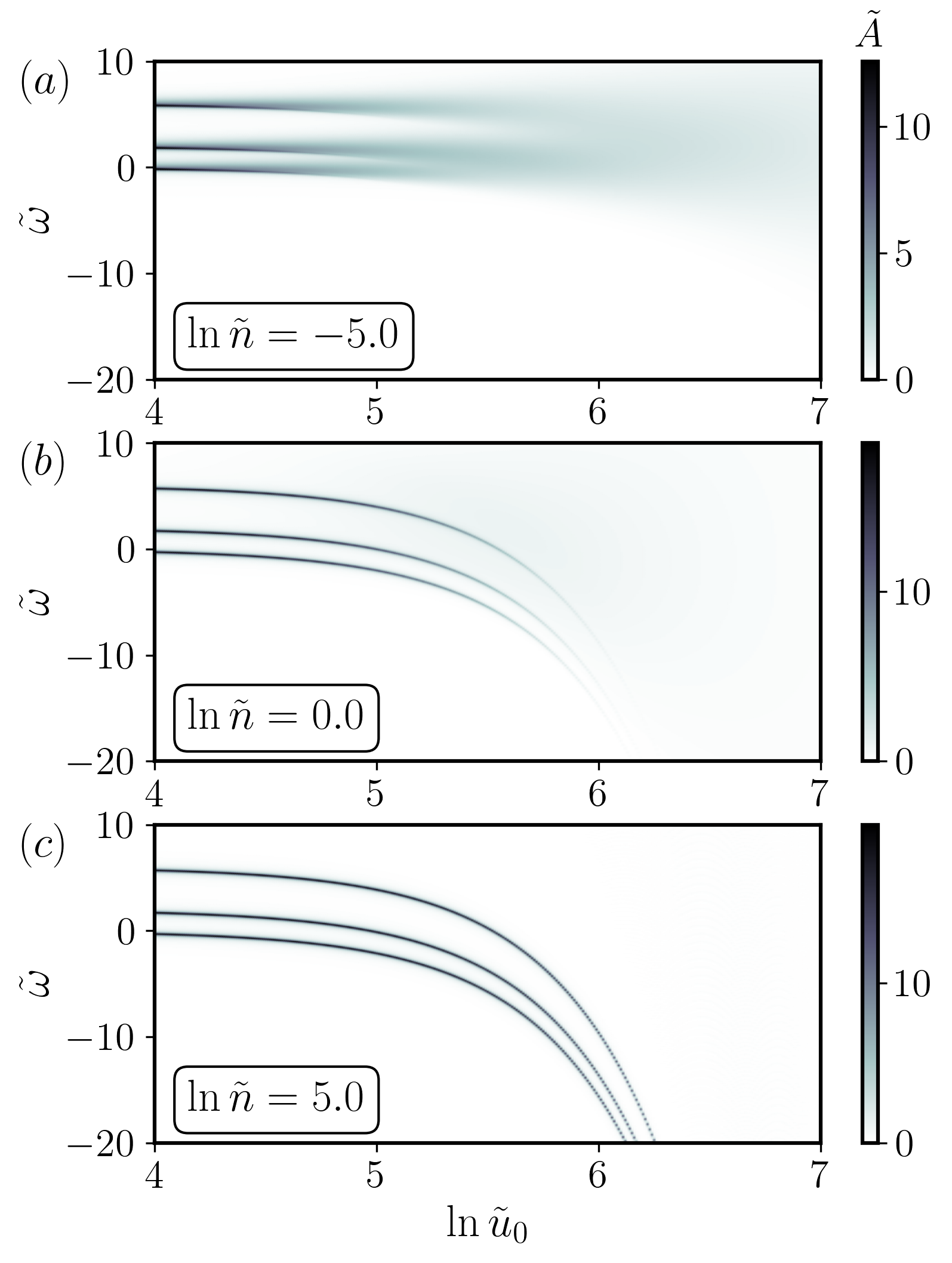}
		\caption{Angulon spectral function in dependence of the rotor-boson interaction magnitude,  $\tilde{u}_{0}$ and $\tilde{u}_{1}=\tilde{u}_{0}/1.75$, in three density regimes.}
		\label{fig:Spectrum_u0}
	\end{figure}

	We show the spectral function in dependence on the rotor-boson interaction amplitude $\tilde{u}_{0}$ in three density regimes in Fig.~\ref{fig:Spectrum_u0}. In all the cases, the spectral peaks are sharp at weak interactions and become unstable when we increase the interaction magnitude. Particularly, in the strong-interaction regime, the spectral lines are significantly broadened due to a large number of phonon excitations in the angulon dressing cloud.

	\section{Rotational Spectroscopy}\label{sec:spectroscopy}
	
	The transition energy between molecular rotational states is experimentally studied using rotational spectroscopy, which involves applying a microwave or a laser field to a molecule trapped e.g.\ in a superfluid nanodroplet. The response can be directly related to the angulon spectral function, Eq.~\eqref{eq.spectrum}, which can be re-expressed in term of a basis of many-body eigenstates, $|f\rangle$, of $\hat{H}$ as:
	\begin{equation}\label{eq:spec}
		\begin{split}A(\omega)= & 2\pi\sum_{f}|\langle f|0\rangle|^{2}\delta(\omega-E_{f})\\
			= & \int_{-\infty}^{\infty}dt\langle0|e^{-i\hat{H}t}|0\rangle e^{i\omega t},
		\end{split}
	\end{equation}
	where here $|0\rangle$ represents the non-interacting state of the combined rotor-bath system. As evident from this expression, the spectral function thus encodes the response of the bath to a sudden switching on of the impurity-bath interaction. Since the spectral function provides a direct measure for how well-defined the quasiparticle is, thereby we will refer to it as the quasiparticle spectrum to avoid misunderstanding.
	
\begin{figure}
		\centering  
		\includegraphics[width=1.0\columnwidth]{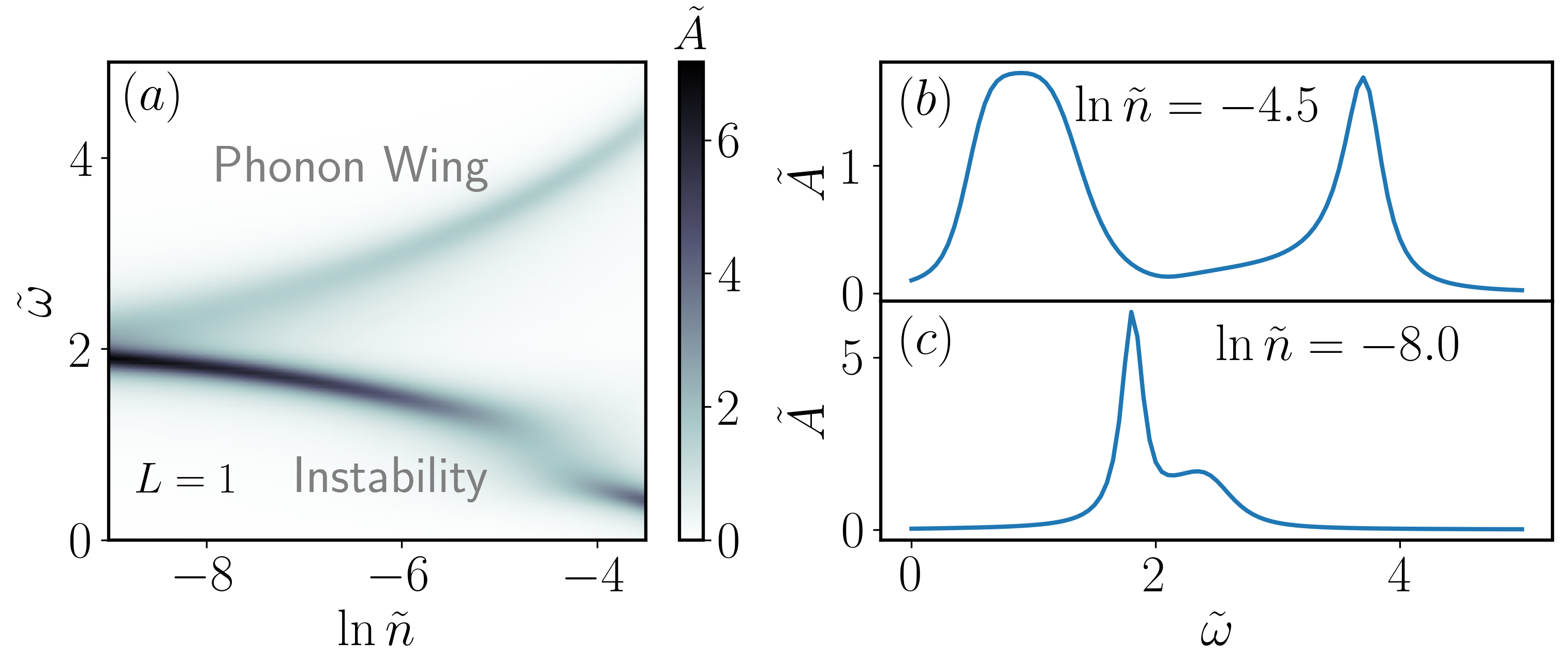}
		\caption{(a) Spectral function for $L=1$ obtained using the single-excitation ansatz. (b), (c) Cuts of the spectral lines at fixed densities.}
		\label{fig:Spectrum_Chevy}
	\end{figure}

	Prior work \cite{schmidt_rotation_2015} with the single-excitation ansatz predicts a regime of angulon instability in the spectrum, which explains the anomalous broadening of spectral lines as can be seen in Fig.~\ref{fig:Spectrum_Chevy} and observed in experiments \cite{morrison_rotational_2013}. However, significant phonon wings, which are predicted to occur at the unperturbed rotational transition frequencies and particularly dominate in the instability regime, have not been observed in experimental rotational spectroscopy. This raises a question about the robustness of the model and the origin of the predicted phonon wings.
	
	To answer this question, it is essential to realize that in chemistry experiments, instead of a rapid injection (or rapidly switching on the interactions, as performed in ultracold polaron experiments \cite{Schmidt_2018}), the molecule is prepared in equilibrium with the nanodroplet. Only then the molecule is excited to higher angular momentum states by a laser pulse. Hence, the initial state of the system within a Fermi’s golden rule description (such as Eq.~\eqref{eq:spec}) has to be chosen with care. Accounting for the initial state preparation the expansion for the transition rate is given by
	\begin{equation}
		\begin{split}R(\omega)= & 2\pi\sum_{f}|\langle f|\hat{V}|i\rangle|^{2}\delta(\omega-E_{f}+E_{i})\\
			= & \int_{-\infty}^{\infty}dt\langle i|\hat{V}e^{-i\hat{H}t}\hat{V}|i\rangle e^{i(\omega+E_{i})t}.
		\end{split}
	\end{equation}
	Here the initial state, $|i\rangle$ with energy $E_{i}$, represents the ground state of the angulon (i.e., a state of total angular momentum $L=0$) instead of the vacuum state in the quasiparticle spectrum in Eq.~\eqref{eq.spectrum}. The action of the laser is represented by the operator $\hat{V}$ which is the amplitude of a harmonic perturbation $\hat{\mathcal{V}}(t)=\hat{V}(e^{i\omega t}+e^{-i\omega t})$. We assume a dipole-field interaction between the molecule and the electric field. Since first-order effects dominate for dipolar molecules, the interaction is given by $-\hat{\boldsymbol{d}}\cdot\boldsymbol{\mathcal{E}}(t)\approx-\mu_{0}\mathcal{E}_{0}\cos\omega t\cos\hat{\theta}$. Here $\mu_{0}$ is the dipole moment of the molecule and $\mathcal{E}_{0}$ is the amplitude of the electric field. Then the laser perturbation reads $\hat{V}=-\mu_{0}\mathcal{E}_{0}\cos\hat{\theta}$, which results in a transition changing rotor angular momentum by one. 
	
	The perturbation $\hat{V}$ only acts on the rotor state. As a result, it does not modify the variational manifold for both the single-excitation and coherent state ansatz which will be illustrated in detail in the following two subsections. In brief, we label the ground state in the $L=0$ channel as $|\psi_{00}\rangle$, where the second index indicates $M$ for the single-excitation ansatz and $n$ for the coherent state ansatz. 
	
	The perturbation can be written as 
	\begin{equation}\label{eq:perturbation}
		\hat{V}\sim\cos\hat{\theta}=\sqrt{\frac{4\pi}{3}}Y_{10}(\hat{\theta}),
	\end{equation}
	which can be expanded in the angular momentum basis. It excites the ground state to the $L=1$ sector, such that $\cos\hat{\theta}|\psi_{00}\rangle\equiv\sqrt{\frac{1}{3}}|\psi_{10}^{\prime}\rangle$, where $|\psi_{10}^{\prime}\rangle$ labels an unnormalized state for both the single-excitation ansatz and coherent state ansatz with the quantum number $L=1$. For convenience, we introduce 
	\begin{equation}
		\begin{split}\tilde{R}(\omega)\equiv & \frac{3}{(\mu_{0}\mathcal{E}_{0})^{2}}R(\omega)\\
			= & \int_{-\infty}^{\infty}dt\langle\psi_{10}^{\prime}(0)|\psi_{10}^{\prime}(t)\rangle e^{i(\omega+E_{0})t},
		\end{split}
	\end{equation}
	where $E_{0}$ indicates the ground state energy. Here the time-evolution is still governed by the same equations of motion and one can numerically calculate the absorption spectrum using real-time evolution. We will next illustrate the calculation for both the coherent state and single-excitation ansatz.

	\subsection{Coherent state ansatz}
	
	We first consider the coherent states ansatz. The ground-state in the laboratory frame is given by $\hat{S}(|000\rangle\otimes|C_{0}\rangle)$.	It is worth mentioning that the $\cos\hat{\theta}$ perturbation is invariant under the rotational transformation acting on bosons, $[\hat{S},\hat{V}]=0$. Then the spectrum can be written as 
	\begin{equation}
		R(\omega)=  \int_{-\infty}^{\infty}dt\langle\psi_{00}|\hat{V}e^{-i\hat{\mathcal{H}}t}\hat{V}|\psi_{00}\rangle e^{i(\omega+E_{0})t}.
	\end{equation}
	The ground-state wavefunction and energy have been found in Eq.~\eqref{eq:coherent_ground_wavefunction} and Eq.~\eqref{eq:coherent_ground_energy}. Next one can apply the perturbation to the ground state: 
	\begin{equation}
		\hat{V}|\psi_{00}\rangle\sim|\psi_{10}^{\prime}(t=0)\rangle=|100\rangle|C_{0}\rangle.
	\end{equation}
	
	Surprisingly, even though the coherent state predicts a rich spectral function in the intermediate-density regime as shown in Fig.~\ref{fig:QuasiparticleSpectrum}, the corresponding rotational transition spectrum $R(\omega)$  is always sharp and trivial.  
	Formally, this is caused by the fact that the mean values of the angular momentum operators $\hat{\boldsymbol{\Lambda}}$ and $\hat{\boldsymbol{J^{\prime}}}$ for the initial state $|100\rangle|C_{0}\rangle$ vanish during the whole time evolution. 
	According to Eq.~\eqref{eq:rte_beta} and Eq.~\eqref{eq:rte_g}, this renders the time evolution trivial, and the coherent state part does not evolve at all. As for the angular state, with the constraint of $\boldsymbol{\Lambda}=0$, the equation of motion reduces to an ordinary linear differential equation:
	\begin{equation}\label{eq:37}
		i\partial_{t}g_{n}=g_{n}[BL(L+1)+\sum_{k\lambda}V_{\lambda}(k)\text{Re}\beta_{k\lambda0}].
	\end{equation}
	Given the initial condition, only the $n=0$ channel evolves up to a phase. As a result, Eq.~\eqref{eq:37} is trivially solved and we find
	\begin{equation}
		\tilde{R}(\omega)= 2\pi \delta(\omega-BL(L+1)).
	\end{equation}
	This demonstrates that despite its power in describing the renormalization of rotational constants, the coherent state approach yields only a trivial rotational spectrum and is thus clearly insufficient to capture the physics of rotational spectroscopy experiments. To remedy this shortcoming it would at least have to be combined with single-particle excitations, allowing the state to evolve among the variational tangent space  \cite{shi_variational_2018}.
	
	\subsection{Single-excitation ansatz}
	
	Next we examine the single-excitation ansatz of Eq.~\eqref{eq:single_excitation_ansatz}. The self-consistent calculation in Ref.~\cite{schmidt_rotation_2015} yields the ground-state variational parameters:
	\begin{equation}\label{eq:se_10}
		\begin{split}Z_{(0)}^{1/2}= & \left[ 1+\sum_{k\lambda}\frac{V_{\lambda}^{2}(k)}{(W_{k\lambda}-E_{0})^{2}}\right] ^{-1},\\
			\beta_{k\lambda\mu}^{(0)}= & (-1)^{\lambda+1}\delta_{\lambda,\mu}\frac{V_{\lambda}(k)}{W_{k\lambda}-E_{0}}Z_{(0)}^{1/2},
		\end{split}
	\end{equation}
	which can be obtained by inserting the ground state energy $E_{0}$. Next, one can evaluate the laser perturbation acting on this state, $\hat{V}|\psi_{00}\rangle\sim|\psi_{10}^{\prime}\rangle$, with
	\begin{equation}\label{eq:42}
		\begin{split}
			|\psi_{10}^{\prime}\rangle&\equiv Z_{(0)}^{1/2}|0\rangle|10\rangle\\
			&+\sum_{k\lambda\mu}\sum_{jm}\beta_{k\lambda j}^{\prime}C_{jm,\lambda\mu}^{10}|k\lambda\mu\rangle|jm\rangle,
		\end{split}
	\end{equation}
	and
	\begin{equation}
		\beta_{k\lambda j}^{\prime}=(-1)^{\lambda}\sqrt{\frac{2j+1}{3}}\beta_{k\lambda\lambda}^{(0)}C_{j0,\lambda0}^{10}.
	\end{equation}
	As evident from Eq.~\eqref{eq:42}, the wavefunction $|\psi_{10}^{\prime}\rangle$ respects the form of the single-excitation ansatz. Hence the equations of motions remain intact:
	\begin{equation}
		\begin{split} & i\partial_{t}Z^{1/2}=BL(L+1)Z^{1/2}+\sum_{k\lambda j}(-1)^{\lambda}V_{\lambda}(k)C_{L0,\lambda0}^{j0}\beta_{k\lambda j},\\
			& i\partial_{t}\beta_{k\lambda j}=W_{kj}\beta_{k\lambda j}+(-1)^{\lambda}V_{\lambda}(k)C_{L0,\lambda0}^{j0}Z^{1/2},
		\end{split}
	\end{equation}
	with initial conditions given by 
	\begin{equation}
		Z^{1/2}(0)=Z_{(0)}^{1/2}; \ \ \ \	\beta_{k\lambda j}(0)= \beta_{k\lambda j}^{\prime}.
	\end{equation}

	\begin{figure}
		\centering 
		\includegraphics[width=1.0\columnwidth]{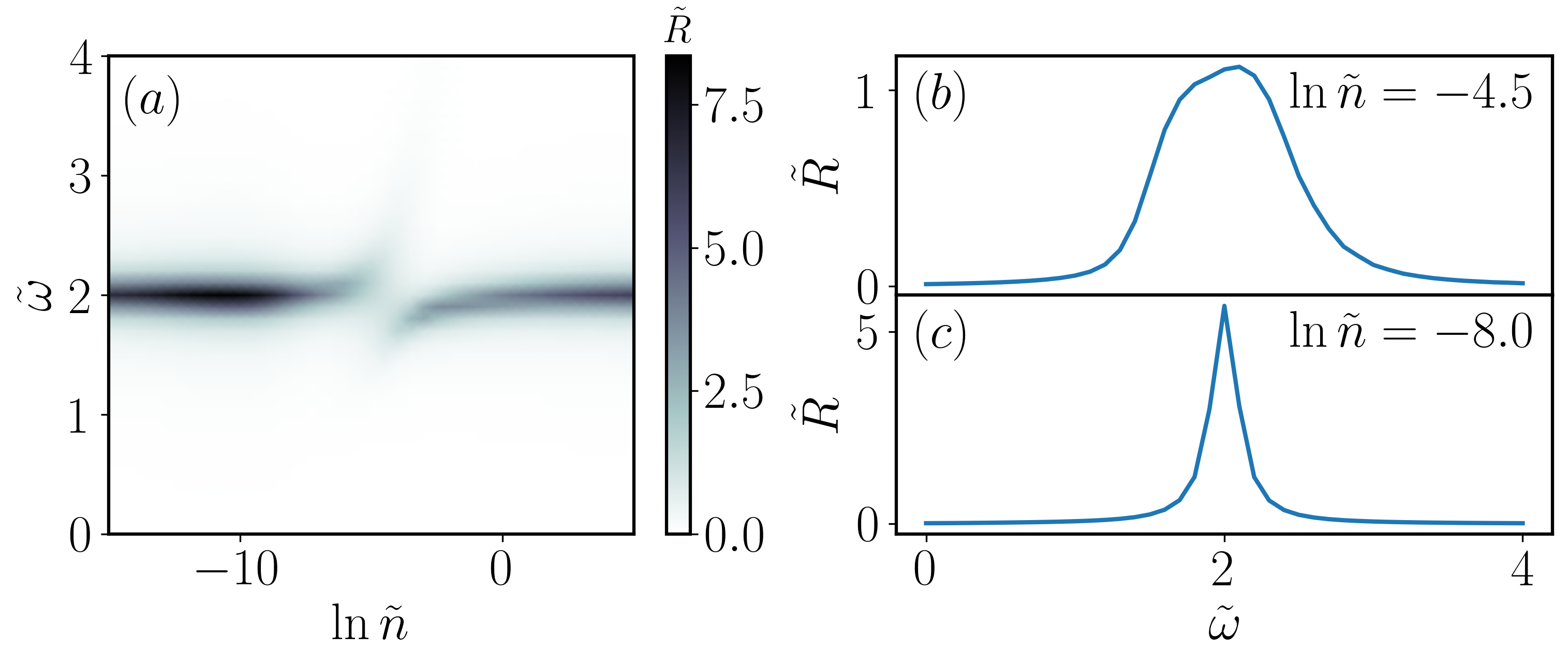}
		\caption{(a) Rotational transition spectrum predicted by the single-excitation ansatz. Here we introduce the dimensionless rotational spectroscopy, $\tilde{R}(\omega)\equiv R(\omega)/3(\mu_{0}\mathcal{E})^{2}$. (b), (c) Cuts of spectral lines obtained by the single-excitation ansatz in the low- and large-density regimes.}
		\label{fig:Spectroscopy}
	\end{figure}

	Fig.~\ref{fig:Spectroscopy} (a) shows that the rotational transition spectrum $R(\omega)$ still prominently features  the instability regime. Fig.~\ref{fig:Spectroscopy} (b) shows a corresponding cut in the intermediate-density regime. In the instability regime, the quasiparticle is unstable and the spectral lines are significantly broadened, as observed in experiments \cite{cherepanov_fingerprints_2017,morrison_rotational_2013}. Remarkably, phonon wings are completely suppressed. Specifically, in contract to the spectral line away from the instability regime is sharp as shown in Fig.~\ref{fig:Spectroscopy} (c). This is in contrast to the quasiparticle spectral function shown in Fig.~\ref{fig:Spectrum_Chevy}, where pronounced phonon wings are visible and the spectral line even splits into two. 
	
	The phonon wing can be understood as arising from excited states of the many-body system, that features `unbound' phonons excited by the molecule's rotation. In rotational spectroscopy, a laser, which only interacts with the rotor, excites the ground state efficiently only to the lower branch of the $L=1$ sector. Since already the initial state is dressed by phonon excitations, it appears that the overlap to such excited states is highly suppressed. This is  in contrast to the spectral function where the overlap between the interacting state and a state without dressing by phonon excitation is relevant.
	In other words, the wavefunction for the upper branch has no overlap with the state $|\psi_{10}^{\prime}\rangle$. 
	
	Thus we make the finding that if the laser perturbation and the equilibrium initial state are correctly taken into account, the spectroscopic response function $R(\omega)$ exhibits an instability regime but no phonon wings, consistent with rotational spectroscopy experiments. The importance of the perturbation is obvious since different types of perturbations correspond to different experimental setups. For example, in electronic excitation spectra, the helium degrees of freedom are also excited and phonon wings are observed \cite{toennies_superfluid_2004,hartmann_direct_1996, doi:10.1063/1.1804945}.
	
	As we have seen, the equilibrium initial state plays a key role in suppressing the phonon wings. This can be demonstrated by performing the calculation for a vacuum initial state:
	\begin{equation}
		\hat{V}|0\rangle|00\rangle	\sim  |0\rangle|10\rangle.
	\end{equation}
	Here $|\psi_{10}^{(0)}\rangle=|0\rangle|10\rangle$ is the vacuum state at $L=1$. In this case, the absorption spectrum is given by
	\begin{equation}
		\begin{split}\tilde{R}(\omega) =\int_{-\infty}^{\infty}dt\langle\psi_{10}^{(0)}|e^{-i\hat{H}t}|\psi_{10}^{(0)}\rangle e^{i(\omega+E_{0})t},\end{split}
	\end{equation}
	which coincides with the angulon spectral function for $L=1$ sector up to a constant energy offset. Hence the phonon wings will still appear. 
	
	Our finding highlights the importance of the equilibrium initial state. It, however, also suggests a way to observe a phonon wing in rotational spectroscopy. While the type of perturbation for rotational spectroscopy is fixed, e.g. Eq.~\eqref{eq:perturbation}, one could consider for instance using aligned molecules as an initial state \cite{PhysRevLett.119.073202} or preparing a non-equilibrium initial state, for example, turning on the laser pulse before the molecule reaches equilibrium with the bath and by performing time resolved pump-probe experiments \cite{doi:10.1126/science.1123904,worner_following_2010}.

	\section{Conclusion and Discussion}\label{sec:conclusion}
	
	In the present article, we variationally studied the angulon model which is an effective description for a rotating molecule immersed in a bosonic bath. We intended to resolve two key issues raised in the previous research:
	\begin{enumerate}
		\item[(1)] the prediction of an apparently non-physical increase of the effective rotational constants in the intermediate coupling regime, and
		\item[(2)] finding an explanation for the unobservable phonon wings in rotational spectroscopy experiments.
	\end{enumerate}
	
	To address the first point we proposed a coherent state ansatz in the co-rotating frame that goes beyond previous approaches by taking into account a macroscopic angulon dressing cloud that is self-consistently determined. In this approach, first, a rotational transformation to the molecular frame is performed, partially decoupling the impurity and the bath in the rotated Hamiltonian. Vice versa, the transformation can also be interpreted as acting on the state where it leads to the entanglement of the impurity and bath. Next, we considered a product state ansatz of the angular and a bosonic coherent state, which are treated fully variationally. By tracing out the angular or bosonic state for such a product state structure, an efficient model can be obtained, enabling effective numerics. Specifically, we optimized the variational parameters in an iterative way. Using this technique, we discovered that the ground state can be described by a product state between an anomalous spin coherent state and a bosonic coherent state. Importantly, this ansatz always predicts a \textit{decreased} effective rotational constant, consistent with experimental findings. Using the real-time evolution of the coherent state ansatz, we then predicted the angulon spectral function. It showed that the number of phonon excitations grows to a large value in the intermediate-density regime while being accompanied by a significant broadening of spectral lines.
	
	In order to address the second challenge, we used Fermi's golden rule to examine the rotational spectroscopy for $L=0\rightarrow1$ transitions. We took into account the laser perturbation and the effect of the formation the interacting equilibrium initial state, neglected in previous studies. As a result, we found that the instability regime predicted by the single-excitation ansatz, where spectral lines are broadened, is robust.	The phonon wings, which are found in the spectral function, cannot be observed using conventional rotational spectroscopy, because the equilibrium initial state has no overlap with the excited states. This resolves the conflict between theory and experiment of observing no phonon wings despite their presence in simple angulon theory.
	
	In the present work, we employed both the single-excitation and the coherent state ansatz. However, they both can only explain different parts of the experimental results. While the coherent state which is the exact solution in the slowly-rotating limit, works well in describing the ground-state properties, such as the effective rotational constants, it is insufficient in describing excited-state properties, such as important for fully frequency-resolved rotational spectroscopy. On the contrary, the single-excitation ansatz works well for describing excited-state properties but performs badly in describing the ground state. 
	Hence, it is evident that further work in generalizing the ansatz is required. 
	One promising direction is to consider a single excitation acting on top of a variational coherent state, going beyond Ref.~\cite{schmidt_deformation_2016}, in which the coherent state variational parameters are fixed.
	
	Moreover, we note that the coherent state approach is challenging in the low-density regime when considering the $L\geq2$ sector. Practically, this originates from the denominator of Eq.~\eqref{eq:selfconsistent}, which can reach a singular point, implying that the self-consistent equations do not guarantee a solution. Specifically, this can be seen from the expression
	\begin{equation}
		\begin{split} & W_{k1}^{2}-4B^{2}(L-\Lambda^{x})^{2}\\
			= & (\omega_{k}^{2}+2B\omega_{k})-4B^{2}[(L-\Lambda^{x})^{2}-1].
		\end{split}
		\label{eq:singular}
	\end{equation}
	When $(\omega_{k}^{2}+2B\omega_{k})\geq0$ and $[(L-\Lambda^{x})^{2}-1]>0$, Eq.~\eqref{eq:singular} can vanish as a function of $k$. Since $\Lambda^{x}\leq L$, it is always negative for $L=0,1$. For $L\geq2$, however, it is only negative when $\Lambda^{x}$ is sufficiently large, which necessitates strong interactions or large density. A similar situation occurs when numerically performing the imaginary-time evolution given by Eq.~\eqref{eq:coherent_ite}, in which the static variational parameter $\beta$ turn out to be unphysical in the low-density regime for the $L\geq2$. It will be interesting to further explore this regime, which may be related to a dynamic instability in the system.
	
	Finally, it is important to realize that the Bogoliubov approximation is employed to derive the angulon model~\cite{lemeshko_molecular_2017}. This restricts us to treat the boson-boson interaction within a mean-field framework, and this approximation is strictly only valid when the interactions are weak and the Bogoliubov phonons are stable. With the coherent state approach, it is, however, also in principle possible to directly deal with the first-principle model, i.e., a quantum rotor immersed in the interacting Bose gas. Further exploration in this direction holds promise to reveal deeper insight into the quasiparticle instability regime as well as into higher angular momentum sectors.

	\acknowledgments
	
	We thank Ignacio Cirac, Christian Schmauder, and Henrik Stapelfeldt for valuable discussions. We acknowledge support by the Max Planck Society and the Deutsche Forschungsgemeinschaft under Germany's Excellence Strategy – EXC-2111 – 390814868. M.L.~acknowledges support by the European Research Council (ERC) Starting Grant No.~801770 (ANGULON). T.S. is supported by National Key Research and Development Program of China (Grant No. 2017YFA0718304), and the	National Natural Science Foundation of China (Grants No. 11974363, No. 12135018, and No. 12047503).

	\appendix

	\section{Anomalous spin coherent state}\label{app:SpinCoherent}
	Here we provide a derivation of the ground state of the effective single anomalous spin model, cf. Eq.~\eqref{eq:H_r}, in terms of the angular momentum basis $|LMn\rangle$.
	The spin-dependent part of the Hamiltonian is given by 
	\begin{equation}
		\hat{H}=-\boldsymbol{n}\cdot\hat{\boldsymbol{J^{\prime}}},
	\end{equation}
	where $\boldsymbol{n}=(\sin\theta\cos\phi,\sin\theta\sin\phi,\cos\theta)$. The anomalous angular momentum operators satisfy the algebra:	
	\begin{align}
		\boldsymbol{\hat{J}}^{\prime2}|LMn\rangle & =L(L+1)|LMn\rangle,\\
		\hat{J}^{\prime z}|LMn\rangle & =n|LMn\rangle,\\
		\hat{J}^{\prime\pm}|LMn\rangle & = C^{\pm}_{Ln}  |LM(n\mp1)\rangle,
	\end{align}
	where $C^{\pm}_{Ln} = \mp\frac{1}{\sqrt{2}}\sqrt{(L\pm n)(L\mp n+1)}$.
	
	Next we introduce the anomalous rotation operators,
	\begin{equation}
		\hat{D}^{\prime}(\alpha,\beta,\gamma)=e^{-i\alpha\hat{J}^{\prime z}}e^{-i\beta\hat{J}^{\prime y}}e^{-i\gamma\hat{J}^{\prime z}},
	\end{equation}
	which, unlike the normal rotation operators, indicate left-handed rotation \cite{morrison_guide_1987,https://doi.org/10.1002/cmr.a.21385}. The Hamiltonian can be diagonalized by the transformation,
	\begin{equation}
		\hat{H}=-\hat{D}^{\prime}(-\phi,-\theta,0)\hat{J}^{\prime z}\hat{D}^{\prime\dagger}(-\phi,-\theta,0).
	\end{equation}
	This makes evident that the ground state can be expressed as
	\begin{equation}
		|\psi_{0}\rangle=\hat{D}^{\prime}(-\phi,-\theta,0)|LML\rangle=\sum_{n}g_{n}|LMn\rangle,
	\end{equation}
	where the superposition coefficients are given by $g_{n}=D_{nL}^{\prime L}(-\phi,-\theta,0)$. Here $D_{nm}^{\prime L}$ are the matrix elements of the anomalous Wigner D-matrix, defined as 
	\begin{equation}
		\begin{split}&D_{nm}^{\prime L}(\alpha,\beta,\gamma)\\ \equiv  & \langle LMn|e^{-i\alpha\hat{J}^{\prime z}}e^{-i\beta\hat{J}^{\prime y}}e^{-i\gamma\hat{J}^{\prime z}}|LMm\rangle,\\
			= & e^{-i\alpha n-i\gamma m}d_{nm}^{\prime L}.
		\end{split}
	\end{equation}
	The small Wigner d-operator and matrix elements are defined as $\hat{d}^{\prime}(\beta)\equiv e^{-i\beta\hat{J}^{\prime y}}$ and $d_{nm}^{\prime L}=\langle LMn|e^{-i\beta\hat{J}^{\prime y}}|LMm\rangle$, respectively.
	
	Following Schwinger's oscillator method \cite{sakurai2011modern,schwinger1965quantum}, we derive an analytical expression for the anomalous d-matrix, which is given by 
	\begin{equation}
		\begin{split}&d_{m^{\prime}m}^{\prime(L)}(\beta)=  \sum_{k}(-1)^{k+m+m^{\prime}}\\
			\times&\frac{\sqrt{(j-m)!(j+m)!(j+m^{\prime})!(j-m^{\prime})!}}{(j-m-k)!k!(j-m^{\prime}-k)!(k+m+m^{\prime})!}\\
			\times &(\cos\frac{\beta}{2})^{2j-2k-m-m^{\prime}}(\sin\frac{\beta}{2})^{2k+m+m^{\prime}}.
		\end{split}
	\end{equation}
	
	In deriving the anomalous spin coherent state, we consider the special case of the Wigner d-matrix, $m=L$. Then, one can directly act with rotation operator on the state, i.e. $\hat{D}^{\prime}(-\phi,-\theta,0)|LML\rangle$, to obtain the superposition coefficients Eq.~\eqref{eq:spinCoherentState}.

	\section{Equations of motion for the single-excitation ansatz}\label{app:ChevyEOM}
	The single-excitation ansatz has been introduced in Eq. \eqref{eq:single_excitation_ansatz}. Its variational energy can be obtained by solving the self-consistent equation \cite{schmidt_rotation_2015}:
	\begin{equation}
		E=BL(L+1)-\Sigma_{L}(E),
	\end{equation}
	where the self-energy is given by
	\begin{equation}
		\Sigma_{L}(E)=\sum_{k\lambda j}\frac{V_{\lambda}^{2}(k)(C_{L0,\lambda0}^{j0})^{2}}{W_{kj}-E}.
	\end{equation}
	The variational parameters are given by 
	\begin{equation}
		|Z^{(0)}|=\left[ 1+\sum_{k\lambda}\frac{V_{\lambda}^{2}(k)}{(W_{k\lambda}-E_{0})^{2}} \right] ^{-1},
	\end{equation}
	and
	\begin{equation}
		\beta_{k\lambda j}^{(0)}=\delta_{\lambda,j}\frac{(-1)^{\lambda+1}V_{\lambda}(k)}{(W_{kj}-E_{0})^{2}}|Z^{(0)}|^{1/2}.
	\end{equation}
	
	The equations of motion of the variational parameters are derived, in turn, based on the time-dependent variational principle:
	\begin{equation}
		\begin{split} & \mathcal{L}=\langle\psi|i\partial_{t}-\hat{H}|\psi\rangle\\
			& \frac{d}{dt}\frac{\partial\mathcal{L}}{\partial\dot{f}}-\frac{\partial\mathcal{L}}{\partial f}=0
		\end{split}
	\end{equation}
	where $f$ denotes the time-dependent variational parameters $Z^{1/2}(t)$ and $\beta_{k\lambda j}(t)$.

	\clearpage
	
	\bibliographystyle{apsrev4-1}
	\bibliography{library}	
\end{document}